\documentclass[journal]{IEEEtran}

\usepackage{color,soul}
\usepackage{subcaption}
\usepackage{amsmath,commath}

\ifCLASSINFOpdf
\else
\fi
\usepackage{array}
\newcolumntype{x}[1]{>{\centering\arraybackslash\hspace{0pt}}p{#1}}
\usepackage{xcolor}
 \usepackage{setspace}
\usepackage{amsmath,amssymb}
\usepackage{graphicx,amssymb}
\usepackage{algorithm,algpseudocode}
\usepackage{xcolor,soul}
\usepackage{amsthm,cite}  
\newtheorem*{assumption*}{\assumptionnumber}
\providecommand{\assumptionnumber}{}

\usepackage{mathrsfs}
\usepackage{caption}
\usepackage{url}
\usepackage{braket}
\usepackage{mathtools} 
\usepackage[utf8]{inputenc}
\usepackage{tablefootnote}
\usepackage{tabularx}

\newcolumntype{L}[1]{>{\raggedright\arraybackslash}p{#1}}
\newcolumntype{C}[1]{>{\centering\arraybackslash}p{#1}}
\newcolumntype{R}[1]{>{\raggedleft\arraybackslash}p{#1}}
\usepackage{textcomp}\usepackage[numbers,sort&compress]{natbib}

\makeatletter
\def\BState{\State\hskip-\ALG@thistlm}
\makeatother
\begin{document}
%
\title{Divergence Framework for EEG based Multiclass Motor Imagery Brain Computer Interface}
\author{Satyam Kumar, Tharun Kumar Reddy, and Laxmidhar Behera \IEEEmembership{Senior Member, IEEE}
\thanks{Satyam Kumar, Tharun Kumar Reddy, and Laxmidhar Behera are with the Department
of Electrical Engineering, Indian Institute of Technology, Kanpur, Kanpur
208016, India (e-mails: satyamk@iitk.ac.in, tharun@iitk.ac.in, lbehera@iitk.ac.in)}}
\maketitle
\begin{abstract}
Similar to most of the real world data, the ubiquitous presence of  non-stationarities in the EEG signals significantly perturb the feature distribution thus deteriorating the performance of Brain Computer Interface. In this letter, a novel method is proposed based on Joint Approximate Diagonalization (JAD) to optimize stationarity  for multiclass motor imagery Brain Computer Interface (BCI) in an information theoretic framework.  Specifically, in the proposed method, we estimate the subspace which optimizes the discriminability between the classes and simultaneously preserve stationarity within the motor imagery classes. We determine the subspace for the proposed approach through optimization using gradient descent on an orthogonal manifold. The performance of the proposed stationarity enforcing algorithm is compared to that of baseline One-Versus-Rest (OVR)-CSP and JAD on publicly available BCI competition IV dataset IIa. Results show that an improvement in average classification accuracies across the subjects over the baseline algorithms and thus essence of  alleviating within session non-stationarities. 
\end{abstract}
%
\IEEEpeerreviewmaketitle
\section{Introduction}
\IEEEPARstart{E}{lectroencephalogram (EEG)}, a type of neural signal (recorded non-invasively from the scalp), is commonly used neuroimaging technique to process and decode the commands in a Brain-Computer Interface (BCI) setup. BCI establishes direct communication between the brain and an external device through signal recordings of brain activity. 
Motor Imagery is one of the most popular paradigms for voluntary control of BCIs, i.e., imagining the movement of right or left hands. 
A generic BCI pipeline includes preprocessing, feature extraction and classification steps. 
The EEG signals recorded suffer from low signal to noise ratio due to volume conduction and non-stationarities. In the preprocessing step, spatial filtering is used to improve the signal to noise ratio and further characterization of motor imagery based task-induced changes. Common Spatial Patterns (CSP) algorithm has been one of the gold standard techniques to perform spatial filtering in a binary class motor imagery \cite{lotte2018review}\cite{lotte2007review}. 
CSP is vulnerable to non-stationarities present in the EEG signal. 
Authors in \cite{von2009finding}\cite{von2010finding} proposed stationary subspace analysis (SSA) to extract stationary components of the EEG signal and further used the extracted stationary components for motor imagery analysis. SSA doesn't use the class specific information and hence is a unsupervised technique. In the paper \cite{samek2012brain}, Samek et al.  proposed a supervised method of optimizing the stationarity in the different groups of the trials with grouping based on the tasks performed by the subject. In the case of motor imagery BCI with two classes, the grouping is two different motor imagery classes. Authors in \cite{samek2012brain} also proposed to optimize the discriminativity and stationarity simultaneously. Later, Samek et al. \cite{samek2016robust}\cite{samek2014divergence} proposed a composite divergence based framework to optimize various types of non-stationarities for CSP to incorporate the stationarity. Horev et al. used a riemannian geometry framework to optimise stationarity in binary class motor imagery BCI \cite{horev2016geometry}. 

Initially, CSP algorithm was explicitly derived for binary class motor imagery BCIs. Later different strategies such as One-Versus-Rest CSP \cite{blankertz2003boosting}, pairwise binary classification followed by voting \cite{muller1999designing}, Information theoretic filter selection criteria \cite{grosse2008multiclass} and Riemannian geometry based techniques \cite{barachant2012multiclass}\cite{yger2017riemannian}  were proposed for feature extraction and classification in multiclass motor imagery based BCIs. The proposed information theoretic approaches to optimise stationarities in earlier research works \cite{samek2016robust}\cite{samek2014divergence}\cite{samek2012stationary} can not be directly used in multiclass BCI settings.
The primary goal of this letter is to develop an information theoretic divergence based framework for multiclass motor imagery BCI. This letter proposes methods to optimize the spatial filters for multiclass motor imagery in an information-theoretic setting.
Furthermore, we propose a novel optimization strategy for divergence based JAD framework, so that the set of spatial filters preserve the stationarity within the session as well as optimize the discriminability between motor imagery classes.  

Rest of the letter is organized as follows.
Section \ref{sec:Multi} discusses multiclass motor imagery classification strategies and presents an alternative approach as joint approximate diagonalization on an orthogonal manifold. Section \ref{sec:Metho}  explains proposed methods and algorithms to optimize the stationarity in a multiclass information-theoretic setting. Section \ref{sec:Evalu} discusses the dataset and evaluation strategies used to compare the proposed algorithms. Section \ref{sec:Results}
presents the results as well as discuss the findings along with the proposed direction for future endeavors. Section \ref{sec:conclus} concludes the paper.

\section{Multi Class Motor Imagery }\label{sec:Multi}  
\par\textbf{One-Versus-Rest strategy:}
In this approach, for multiclass classification, several binary classification problems are formulated, and CSP algorithm is used for feature extraction of each binary classification.  Similar to CSP algorithm, the OVR approach computes CSP filters that discriminate each class from all the other classes. For each binary classification, generally two spatial filters are selected according to $\alpha$ sorting criteria (sorting according to discriminativity value, check \cite{samek2016robust} for detail) Assume that the class covariance matrix for class $c_i$ is denoted by $\Sigma_{c_i}$ then $\Sigma_{ovr_{i}}$ is defined as follows:
\begin{equation}
    \Sigma_{ovr_i}=\sum_{j\neq i}^{K}p_{j}\Sigma_{c_j}
    \label{equation:OVRCovariance}
\end{equation}
\begin{equation}
    \Sigma_1\mathbf{W}_{c_j}^i=\lambda\Sigma_2\mathbf{W}_{c_j}^i \:\: i\in\{1,2\}, j\in\{1,2,..,K\}
    \label{eq:OVRCSP}
\end{equation}

where $p_{i}$ is fraction of numbers of samples in use for class $c_{i}$ and $K$ is number of distinct motor imagery classes. For class $c_i$ the spatial filters can be calculated using eigenvalue problem (\ref{eq:OVRCSP}) with $\Sigma_1$ substituted by $\Sigma_{c_i}$ and $\Sigma_2$ by $\Sigma_{ovr_i}$. After selection of spatial filters for each class, a consolidated filter matrix can be created as follows:
\begin{equation}
    W=[ W_{c_1}^1, W_{c_1}^2, \cdots , W_{c_K}^1,W_{c_K}^2]
    \label{eq:ConsolidatedW}
\end{equation} 
In (\ref{eq:ConsolidatedW}), each column represents a spatial filter. Once the spatial filter matrix $W$ is obtained, we use it to extract the power features and train a discriminant classifier for classification of trials.

\par\textbf{Joint Approximate Diagonalization: }
CSP by Joint approximate diagonalization (JAD) algorithm is a popular alternative to OVR-CSP for classification of multiclass motor imagery. Given an EEG data of $K$ different classes, CSP by JAD finds a linear transformation $W \in \mathbb{R}^{M\times M}$ that diagonalizes the class covariance matrices $\Sigma_{c_i}$,
\begin{equation}
    W^T\Sigma_{c_i}W=D_{c_i}, \quad i=1, \cdots ,K\:\: \text{and} \:\:\Sigma_{i=1}^{N}D_{c_i}=\mathbf{I}
    \label{eq:JAD}
\end{equation}
with $D_{c_i} \in \mathbb{R}^{M\times M}$ representing diagonal matrices and $I$ is an ${M\times M}$ identity matrix. We refer to CSP by JAD algorithm with JAD for brevity.

\par\textbf{Information theoretic JAD (IT-JAD):}
Gouy Paeiller et al.\cite{gouy2010nonstationary} presented an information theoretic interpretation of (\ref{eq:JAD}). In JAD framework, we estimate the linear transform that jointly diagonalizes the covariance matrices. It can be interpreted as minimizing the KL divergences between set of transformed covariance matrix and a diagonal matirx. The mathematical formulation of IT-JAD can be written as 
\begin{equation}
    F(V)=\sum\limits_{c=1}^K p_{c}D_{kl}(V^T\Sigma_{c}V \parallel diag(V^T\Sigma_{c}V))
    \label{eq:IT-JAD}
\end{equation}
\begin{equation}
    V^*=\underset{V}{\arg\min}\ F(V) 
    \label{eq: JADFNot}
\end{equation}

where $K$ is number of different motor imagery classes (the number of different matrices to be jointly diagonalized). $p_c$ is the prior probability corresponding to class $c$. The transform $V$ can be decomposed as a product of orthogonal matrix and a whitening transform $V^T=RW$ \cite{samek2016robust}. We can write (\ref{eq:IT-JAD}) in terms of orthogonal matrix $R$ as follows:  
\begin{equation}
    J(R)=\sum\limits_{c=1}^K p_{c}D_{kl}(R\bar\Sigma_{c}R^{T} \parallel diag(R\bar\Sigma_{c}R^{T}))
    \label{eq:IT-JAD_modif}
\end{equation}
\begin{equation*}
\text{such that} \quad \bar\Sigma_c=W\Sigma_c W^T
\end{equation*}
\begin{equation}
\sum\limits_{c=1}^K W\Sigma_{c}W^T=I
\end{equation}

The goal is to minimize $J(R)$ so that $R\bar\Sigma_{c}R^T$ is as diagonal as possible.

Consider $\Tilde{J_1}(R)$ be the first term in (\ref{eq:IT-JAD}) and $p_{1}$ be some constant, we will assume its value to be 1.
\begin{align*}
       \Tilde{J_1}(R)&=D_{kl}(R\bar\Sigma_1 R^T \parallel diag(R\bar\Sigma_1 R^T))\\
       &=(log(det({R\bar\Sigma_1 R^T})^{-1}diag(R\bar\Sigma_1 R^T)))\\
       &+\underbrace{tr(diag(R\bar\Sigma_1 R^T)^{-1}{R\bar\Sigma_1 R^T})}_\text{constant}\\
       &=-log(det({R\bar\Sigma_1 R^T})+log(det(diag(R\bar\Sigma_1 R^T))) +C\\
\end{align*}

The gradient of $\Tilde{J_1}(R)$ w.r.t square matrix R can be written from \cite{petersen2008matrix} as 
\begin{equation}
    \nabla_{R}\Tilde J_{1}(R)=-2R^{-1}+2{diag(R\Tilde \Sigma_1 R^T)}^{-1}R\Tilde\Sigma_1
    \label{eq:Derivative}
\end{equation}
The gradient value from  \eqref{eq:Derivative} can be used to calculate the gradient of \eqref{eq:IT-JAD_modif}. The consolidated gradient is further used to optimize $R$ on an orthogonal manifold. Once the value of $J(R)$ is optimized (convergence criteria satisfied), we  calculate the spatial filters using $V^T=RW$.  Next, the spatial filters are sorted (i.e. column vectors of $V$)  according to mutual information filter selection criteria \cite{grosse2008multiclass}. Generally, we select best 8 filters after the sorting, which are further used for multiclass motor imagery classification.

\begin{algorithm}
\caption{DivJAD ($K$ motor imagery classes) }\label{alg:DivJAD}
\hspace*{\algorithmicindent} \textbf{function:} DivJAD($\{\Sigma_1,\Sigma_2,\cdots,\Sigma_K\}$,$d$)\\
 \hspace*{\algorithmicindent} \textbf{Output:} $V$ (Spatial Filter Matrix)
\begin{algorithmic}[1]

\State Estimate whitening transform W=${(\Sigma_1+\cdots+\Sigma_K})^{-\frac{1}{2}}$
\State Initialize random orthogonal rotation matrix $R_{o}$
\State Apply whitening transform     $\Sigma_i=R_{o}\text{W}\Sigma_{i}(R_{o}\text{W})^T$
\While{Iter=maxIter} \Comment{Convergence criteria}
\State Estimate the gradient at Identity on the  manifold 
\State Estimate optimal step size ($U$) using line search 
\State Update $R_{o}$:  $R_{k+1}=UR_{k}$
\State Rotate the data $\Sigma_i=U\Sigma_{i}U^T$
\EndWhile\label{euclidendwhile}
\State Estimate: $V^T=R_{k+1}W$
\State Sort the spatial filters (use ITFE Criteria\cite{grosse2008multiclass})
\State $V=V([:,1:d])$ \Comment{First $d$ columns of the  matrix $V$}

\end{algorithmic}
\end{algorithm}


\section{Methodology}\label{sec:Metho}
We introduced stationarity in multiclass BCI through divergence based framework using two different approaches as follows:
\subsection{DivOVR-CSP-WS}
The regularization term as described in (10) in (\cite{samek2014divergence}) is extended to multiclass in a way that stationarity is preserved in each of the motor imagery classes separately. It can be written as
\begin{equation}
    G(V)=\frac{1}{{K}N}\sum\limits_{c=1}^{K}\sum\limits_{i=1}^N {D}_{kl}(V^T\Sigma_{i,c}V  \parallel  V^T\Sigma_{c}V)
    \label{eq:Statio}
\end{equation}
Where $K$ is number of different classes, $N$ is number of trials in each class and $\Sigma_{i,c}$ is the covariance matrix of $i^{th}$ trial of class $c$. Note that, we assume the number of trials corresponding to each class to be equal. For each OVR model, we incorporate this regularization term separately i.e.
\begin{equation}
    W_i^*=\underset{V}{\arg\max}\ (1-\lambda)F_{i}^{ovr}(V) - \lambda
    G(V)
    \label{eq: gSSA}
\end{equation}
We estimate the spatial filter for each OVR model by optimizing (\ref{eq: gSSA}) using the subspace optimization method discussed in \cite{samek2014divergence}. $\lambda$ is the regularization parameter. The consolidated filter matrix is then constructed from each OVR model. 

\subsection{DivJAD-WS}
In this framework, we incorporate within session stationarity formulated in (\ref{eq:Statio}) together with information theoretic formulation of JAD as follows 
\begin{align}
        \Delta(V)&=\bigg[(1-\lambda)\bigg(\sum\limits_{c=1}^K p_{c}D_{kl}(V^T\Sigma_{c}V \parallel diag(V^T\Sigma_{c}V))\bigg)\nonumber\\
        &+\lambda\bigg(\frac{1}{{K}N}\sum\limits_{c=1}^{K}\sum\limits_{i=1}^N {D}_{kl}(V^T\Sigma_{i,c}V  \parallel  V^T\Sigma_{c}V)\bigg)\bigg]
    \label{eq:CompositeStatioJAD}
\end{align}
Unlike DivCSP-WS formulation as described by (10) in (\cite{samek2014divergence}), the regularization term is ``added", because we are in totality minimizing both, the diagonalization term as well as group stationarity term. The spatial filters $V^*$ are estimated such that 
\begin{equation}
    V^*=\underset{V}{\arg\min}\ \Delta(V)
    \label{eq:Vsymbol}
\end{equation}
we optimize (\ref{eq:Vsymbol}) on the subspace $V$ 
 \begin{equation}
     R^*=\underset{R}{\arg\min}\ (1-\lambda)J(R)+\lambda J_{s}(I_{d}R)
     \label{eq:CompositeDelta}
\end{equation}
Here $J$ and $J_{s}$ represent the diagonalization objective and stationarity objective in terms of $R$ (Orthogonal matrix). In (\ref{eq:CompositeDelta}), we optimize the complete subspace for joint approximate diagonalization term. However, for optimization of stationarity, we use $I_{d}R$ instead of $R$. In layman terms ``$I_{d}R$ " implies to selecting first $d$ rows of the orthogonal matrix $R$ hence selecting first $d$ columns of filter matrix $V$ to enforce stationarity. Using the proposed method, we achieved both the objectives, joint diagonalization of the matrices as well as enforcing stationarity on first $d$ components of the transform.

\begin{table*}[h]

  \centering
      \caption{Cross-validation and testing Classification accuracies (in \%) for each subject calculated using different methods. Optimum $\lambda$ value is selected from the range $[0,1]$ using cross validation}
 \begin{tabular}{lccccccccccc}
\hline
\multicolumn{1}{c}{} & 
\begin{tabular}[c]{@{}c@{}}S1\end{tabular}
& \begin{tabular}[c]{@{}c@{}}S2 \end{tabular}
& \begin{tabular}[c]{@{}c@{}}S3 \end{tabular}
& \begin{tabular}[c]{@{}c@{}}S4 \end{tabular}
& \begin{tabular}[c]{@{}c@{}}S5 \end{tabular}
& \begin{tabular}[c]{@{}c@{}}S6 \end{tabular}
& \begin{tabular}[c]{@{}c@{}}S7 \end{tabular}
& \begin{tabular}[c]{@{}c@{}}S8 \end{tabular}
& \begin{tabular}[c]{@{}c@{}}S9 \end{tabular}
& \begin{tabular}[c]{@{}c@{}}Mean CV \end{tabular}
& \begin{tabular}[c]{@{}c@{}}Mean test \end{tabular}
\\ \hline
     
JAD& 78.51&56.13&84.58&54.82&39.05&47.74&78.63&80.83&73.10 & 65.93 &63.46\\

OVR& 71.01&55.54&82.68&55.24&36.79&47.26&67.08&84.05&73.87 & 63.72 & 62.85\\  

DivJAD-WS& 80.65&58.21&85.60&56.07&39.29&50.00&78.39&84.58&75.77 & 67.62 & 65.16\\

DivOVR-WS&75.65&56.85&85.65&55.60&37.08&49.82&68.87&84.76&75.00 & 65.48 & 64.08\\ 



\hline
  \end{tabular}

 \label{tab:CVresult_all_methods}
\end{table*}
\section{Evaluation}\label{sec:Evalu}
We evaluate the proposed methods on BCI competition IV (dataset IIa)\cite{brunner2008bci}. The paradigm consisted of four different motor imagery classes, namely the movement imagination of the left hand, right hand, feet, and tongue. A total of 288 trials for each subject (9 different subjects) was recorded for training and an equal number of trials for testing phase (session 2). This dataset is balanced with an equal number of trials for all the classes.  A detailed description of the dataset is available on the competition website\footnote{{http://www.bbci.de/competition/iv/}}.

The performance of all the frameworks in this study is assessed using cross-validation accuracy on training set and classification accuracy on the testing set. For the analysis of EEG signals, we extract a time window from 0.5s-3.5s, after motor imagery cue is presented. The signals are further bandpass filtered in a wide band of 8-30 Hz.  In the cross-validation strategy, the training set is randomly chosen as 80\% of the total number of available trials (in training set) for each class (58 for each class and a total of 232) and the remaining data is chosen as the test set. The classification performance on testing fold is evaluated, and the process is repeated 30 times (1x30 fold). The cross-validation performance for each framework is the classification performance averaged across all the 30 folds. 

All frameworks which enforce stationarity incorporate a regularization parameter $\lambda$. The value of $\lambda$ which maximizes cross-validation performance on the training set is the optimum regularization parameter. The optimum regularization parameter is then used to evaluate the performance on the testing set. The regularization parameter $\lambda$ is selected from the set $[0, 0.1, 0.2,\cdots, 1]$.

In OVR-CSP for multiclass, two spatial filters from each OVR case is selected, according to the pre-specified criteria (check   \cite{samek2016robust} for detail). Thus, for this dataset, we use a total of $8$ spatial filters. 
In the JAD framework, an Information theoretic feature extraction technique is used to rank the spatial filters \cite{grosse2008multiclass}. We select $8$ best spatial filters to establish a fair comparison with OVR-CSP frameworks. All the implementations in this paper use shrinkage based estimator for estimation of sample co-variance matrices \cite{ledoit2004well}.

\section{Results and Discussion}\label{sec:Results}
Information theoretic JAD is used for calculation of spatial filters in multiclass motor imagery. Supplementary material compares the classification accuracies of IT-JAD and JAD method. The average classification accuracies across all the subjects are similar for both ITJAD and JAD frameworks. In the proposed approach by Samek et al. \cite{samek2014divergence}, both Information theoretic CSP (based on symmetric KL divergence and without regularization) and CSP yield the same classification accuracies across the subjects. These observations further validate our method and its subsequent use to enforce stationarity in multiclass BCI scenario.  Hence, a divergence based framework can be used to incorporate the stationarity term in an information theoretic formulation.

Table \ref{tab:CVresult_all_methods} summarizes the subject wise average cross-validation (CV) performance across all the folds on the BCI competition IV dataset IIa. The CV accuracies reported for divergence based approaches are the best CV accuracies for different values of $\lambda$ (regularization parameter). Both the stationarity enforcing approaches (DivOVR-WS and DivJAD-WS) outperform the baseline JAD, and OVR approaches in terms of mean classification accuracies across the subjects. Also, the proposed DivJAD-WS outperformed all the baseline methods with an increase in 3.90\% compared to OVR approach and 1.69\% compared to Cardoso based joint diagonalization method for estimating spatial filters. 

Last column of Table \ref{tab:CVresult_all_methods}  presents the classification performance on the testing set for all the methods described in this paper. Similar to that of cross-validation performance, both DivJAD-WS and DivOVR-WS methods outperform their corresponding baseline of OVR and JAD, when mean classification accuracy across the subjects is compared. Also, DivJAD method outperforms both the baseline, i.e., JAD and OVR by 2.31\% and 1.70\% respectively. 

One of the classic shortcomings of classification models in machine learning is over-fitting. Generally, algorithms tend to perform exceptionally well during the cross-validation and in contrast poorly on the unseen test-set. In this paper, for all of the proposed and baseline algorithms, the difference between cross-validation performance and performance on the test-set is quite low, i.e., around $~ 1.5\%-3\%$. These values further validate our efficient cross-validation procedure. Furthermore, our proposed algorithms would be extremely effective in real life BCIs as the parameters selected for a training session (through cross-validation) can be efficiently used for controlling the BCI in a separate session. Lotte and Jeunet \cite{lotte2018defining} proposed run wise cross validation as an efficient metric for offline classification algorithms. The cross-validation strategy used in this paper is similar to the metric proposed in \cite{lotte2018defining} and hence, further reinforces the effectiveness of our proposed framework. The optimum regularization parameter $\lambda$ is data dependent and subsequently subject dependent. Also, the value of the optimum $\lambda$ significantly depends on the set of values of $\lambda$ used for cross-validation. Thus, a further optimization of classification performance can be done by efficiently selecting the regularization set for selection of optimum $\lambda$.

The computational complexity of an algorithm has always been an important characteristic. In this paper, the Divergence based algorithms are dependent on optimization using gradient descent algorithm on the orthogonal manifold. Thus the convergence using gradient descent heavily depends on the initialization of the orthogonal matrix on the manifold. In the proposed DivOVR-WS framework, if the initialization of orthogonal matrix is the corresponding orthogonal matrix from the OVR-CSP solution, the convergence is a lot faster compared to the case when the orthogonal matrix is randomly initialized. Initializing the Orthogonal matrices from the solution of JAD (estimated using \cite{cardoso1996jacobi}) improves the convergence time of DivJAD-WS, but no significant difference in classification accuracy is observed compared to Identity initialization. Also, the convergence of DivOVR-WS is much faster in comparison to DivJAD-WS when initialized from the OVR-CSP solution in contrast to that of Identity initialization of DivJAD-WS. Furthermore, if the initialization for both DivOVR-WS and DivJAD-WS framework is same (other than OVR solution) the computational time for estimating subspaces using DivOVR-WS is increased as comparison to DivJAD-WS  , as it solves the optimization for '$K$' different OVR scenario ('$K$' motor imagery classes) compared to a single composite JAD optimization. However, once the spatial filters are calculated (training phase), the time taken to eventually classify the motor imagery trials (testing phase) remain the same across all four algorithms.

Recently, Riemannian based approaches for classification of motor imagery trials has attracted a lot of interest from the BCI community \cite{barachant2012multiclass}. The classification performance of the proposed methods could further be improved by using the Karcher mean as a reference class covariance matrix instead of the Euclidean mean as the karcher mean has been proven to be better estimate of average covaraince matrix in BCI applications \cite{barachant2012multiclass}\cite{barachant2010riemannian}. Zanini et al. \cite{zanini2018transfer} proposed a  framework to optimize the classification performance based on an Affine transformation strategy on riemannian manifolds. This method tracks the drifts occuring between sessions. In our proposed approach, we optimize stationarity within session, it will be interesting to see the effect of the using affine transform based strategy of \cite{zanini2018transfer} together with our proposed approach to incorporate both within and between session changes.

\section{Conclusion}\label{sec:conclus}
In this letter we proposed two different supervised learning frameworks , called DivJAD-WS and DivOVR-WS, to tackle the problem of non-stationarity present during a session in multiclass motor imagery  BCI. The stationarity was optimized by proposing the novel objective function incorporating the discriminatory and stationarity term. Furthermore, we also introduced a novel method to estimate a subspace which optimizes the discriminability between the classes and parallelly preserving the stationarity within the session in multiclass BCI. The experimental results on BCI competition IV dataset IIa demonstrated that proposed algorithms outperformed their corresponding baseline (DivJAD-WS vs. JAD \& DivOV-WS vs. OVR) on average classification accuracies compared on cross-validation and testing set. Moreover, We also demonstrated the effectiveness of Information theoretic JAD to estimate the spatial filters for multiclass motor imagery BCIs. 


\bibliographystyle{IEEEtran}
\bibliography{refs}

\end{document}